\documentclass{ws-ijmpa}

\begin{document}

\markboth
{Serena Gambino}
{{Bondi and Novikov-Thorne accretion in regular black holes and Simpson-Visser spacetimes}}
\catchline{}{}{}{}{}

\title{Bondi and Novikov-Thorne accretion in regular black holes and Simpson-Visser spacetimes}

\author{Serena Gambino}

\address{Scuola Superiore Meridionale, Largo San Marcellino 10,
Naples, 80138, Italy.\\
Istituto Nazionale di Fisica Nucleare, Sezione di Napoli, Strada Comunale Cinthia,
Naples, 80126, Italy.\\
s.gambino@ssmeridionale.it}

\maketitle

\begin{abstract}
We compare the Bondi spherical accretion model and the Novikov-Thorne thin disc formalism around regular black holes and Simpson-Visser spacetimes, using several equations of state for the accreting fluids. The Bondi model is significantly more sensitive to spacetime geometry and the equation of state, making it more effective at distinguishing between regular and classical solutions. For the Simpson-Visser solutions, however, increasing the regularisation parameter, $\ell$, shifts critical point positions both inward and outward. However, the charged Simpson-Visser extension, modeled by the charge $Q$, does not produce an effect comparable to that of $\ell$.

\keywords{regular black holes; Bondi accretion; Novikov-Thorne accretion; Simpson-Visser spacetime; accretion disc luminosity}
\end{abstract}

\section{Introduction}

Regular black hole (RBH) solutions\cite{Bardeen:1968,Dymnikova:2004,Hayward:2005,Fan-Wang:2016} are able to circumvent the Penrose-Hawking singularity theorems\cite{Hawking:1970,Hawking:1996} allowing finite geometric invariants everywhere while still maintaining an event horizon. The observational signatures of RBHs, such as quasi-periodic oscillations, quasi-normal modes and shadows, have been studied extensively\cite{Boshkayev:2023}.

Using accretion models as an observational tool allows us to test various geometries of spacetime. Bondi spherical accretion model\cite{Bondi:1952,Michel:1972} and the Novikov-Thorne (NT) thin-disc formalism\cite{NovikovThorne:1973} can be used to calculate the mass accretion rates and luminosity profiles of the accretion disc\cite{Boshkayev:2021chc,Boshkayev:2023fft}.The former describes spherically symmetric, pressure-driven accretion flows, while the latter applies to geometrically thin, optically thick accretion discs with angular momentum. Previous work on the accretion of dark matter around compact objects has shown a strong sensitivity to the matter model, particularly for NT accretion see\cite{Boshkayev:2020,Boshkayev:2022,Kurmanov:2021,Kurmanov:2024hpn}.The Simpson-Visser (SV)\cite{Simpson:2018} metric, on the other hand, is characterised by the regularisation parameter $\ell$ and spans geometries from BHs to traversable wormholes, including RBHs and extremal configurations. This makes it a useful framework to investigate how deviations from classical solutions affect observable quantities, such as accretion luminosity.

This proceeding summarises the results of \refcite{Capozziello:2025} (RBHs) and \refcite{Gambino:2025} (SV spacetimes), to which we refer for all derivations and full parameter surveys.

The article is structured as follows: Section \ref{sec:theory} presents the theoretical framework of the two accretion models, the different equations of state (EoS) for fluids, the RBH solutions and the SV spacetime. Section \ref{sec:results} discusses the complete numerical results, and Section \ref{sec:conc} gives the conclusions.

Throughout the text, units are adopted where $G = c = 1$.
\section{Accretion models and spacetimes}\label{sec:theory}
Below, we briefly outline the two growth models and the main equations used in this study. We then present all the solutions adopted.
\subsection{Bondi and Novikov-Thorne accretion models}

We consider the spherically symmetric metric 
\begin{equation}
    ds^{2}=-f(r)dt^{2}+\frac{1}{g(r)}dr^{2}+r^{2}(d\theta^2 + \sin^2\theta d\phi^{2})\,,
\end{equation} 
with $f(r)=g(r)$ for all solutions studied. In terms of spherically symmetric accretion, such as in the Bondi model for a perfect fluid, the conservation of energy-momentum and mass flux yields the following sonic-point equation\cite{Babichev:2004,Michel:1972}
\begin{equation}
  \left[ V^2 - \frac{u^2}{u^2 + g(r)} \right] \frac{du}{u} + 
   \left[ (V^2-1)\frac{f'(r)}{2f(r)} + \frac{2V^2}{r} - \frac{g'(r)}{2(u^2+g(r))} \right] dr = 0\,,
\label{eq:sonic_eq}
\end{equation}
where $V^{2}=d\ln(P+\rho)/d\ln\rho-1$. Then, the critical (sonic) point conditions can be found as follows
\begin{equation}
V_{c}^{2} = \frac{1}{1+4f(r_{c})/[f'(r_{c})r_{c}]},
\qquad
u_{c}^{2} = \frac{g(r_{c})f'(r_{c})r_{c}}{4f(r_{c})}\,.
\label{eq:critical}
\end{equation}
The Bondi mass accretion rate and disc luminosity depend on the adopted fluid solution and are defined as follows: $\dot{M}=-\int d\theta\,d\phi\sqrt{-g}\,T^r_0$ and $L=\eta_{\rm eff}\dot{M}$, where $\eta_{\rm eff}\simeq0.1$\cite{Frank:2002}.

For the NT model, the radiative flux in the equatorial plane is given by the following standard formula
\begin{equation}
F(r)=\frac{\dot{M}}{4\pi\sqrt{-g}}\,\frac{d\Omega/dr}{(E-\Omega L)^{2}}
\int_{r_{i}}^{r}(E-\Omega L)\frac{dL}{d\tilde{r}}\,d\tilde{r}\,.
\label{eq:flux_NT}
\end{equation}
Since the flux is not directly observable, we will use the differential luminosity at infinity to compare the two accretion models. This is given by $r\,d\mathcal{L}_{\infty}/dr=4\pi r\sqrt{-g}\,F(r)$.

\subsection{Fluid models}

\noindent\textbf{Dark fluid.} A specific case of a barotropic fluid, where $P=w\rho=\mathrm{const}$ implies $\mathcal{C}_{s}^{2}=0$, so $V^{2}$ changes sign at $r=r_{\rm crit}$, requiring the redefinition $V^{2}\to-V^{2}$ for $r>r_{\rm crit}$. Therefore, there is no classical critical point analysis, but by redefining the variable $V^2$, we obtain a physical solution for our observables. The velocity and density profiles, for this fluid, read
\begin{equation}
u(r)=-\sqrt{\frac{g(r)}{f(r)}C_{2}-g(r)},\qquad
\rho=-\frac{C_{3}}{r^{2}\sqrt{g(r)/f(r)}\,\sqrt{(g(r)/f(r))C_{2}-g(r)}}\,.
\end{equation}

\noindent\textbf{Exponential (Sofue) profile\cite{Sofue:2013a,Sofue:2013b}.} The density profile reads $\rho(r)=\rho_{0}e^{-r/r_{0}}$, which gives a non-zero value of $\mathcal{C}_{s}^{2}$ and well-defined sonic points for all RBHs, but not for Schwarzschild or Schwarzschild-de Sitter. 

For the SV analysis\cite{Gambino:2025}, we decided to replace the dark fluid with a generic barotropic fluid, i.e. $P(r)=w\rho(r)$, while keeping the exponential profile as second case.

\subsection{Spacetimes geometries}

\noindent\textbf{RBHs\cite{Capozziello:2025}.} We present the explicit form of the metric function for each RBH solution. Hayward solution: $f(r)=1-\frac{2Mr^{2}}{r^{3}+2a^{2}}$; Bardeen solution: $f(r)=1-\frac{2Mr^{2}}{(r^{2}+q_{B}^{2})^{\frac{3}{2}}}$; Dymnikova solution: $f(r)=1-\frac{4M}{\pi r} \bigg[\arctan\bigg(\frac{r}{r_{D}}\bigg)-\frac{rr_{D}}{(r^{2}+r_{D}^{2})}\bigg]$, where $r_D=\pi \frac{q_D^2}{8M}$; Fan-Wang solution: $f=1-\frac{2Mr^{2}}{(r+l_{\rm FW})^{3}}$, with $l_{\rm FW}\leq8/27$. For comparison, we also include the Schwarzschild and Schwarzschild-de Sitter solutions, for which the metric functions are respectively $f(r)=1-\frac{2M}{r}$ and $f(r)=1-\frac{2M}{r}-\frac{\Lambda}{3} r^{2}$.

\noindent\textbf{SV spacetime\cite{SimpsonVisser:2018}.} This metric requires the substitution $r\to\sqrt{x^2+\ell^2}$, which gives the following line element
\begin{equation}
ds^{2}=-\!\left(1-\frac{2M}{\sqrt{x^{2}+\ell^{2}}}\right)\!dt^{2}
+\frac{dx^{2}}{1-2M/\sqrt{x^{2}+\ell^{2}}}
+(x^{2}+\ell^{2})\,d\Omega^{2}\,.
\end{equation}
For $\ell=0$ we have a Schwarzschild solution; for $0<\ell<2M$ a RBH; for $\ell=2M$ an extremal BH and for $\ell>2M$ a traversable wormhole solution with a throat at $x=0$. The sonic-point conditions become\cite{Gambino:2025}
\begin{equation}
V_{c}^{2}=\frac{M}{2\sqrt{x_{c}^{2}+\ell^{2}}-3M},\quad
u_{c}^{2}=\frac{M}{2\sqrt{x_{c}^{2}+\ell^{2}}}\,.
\end{equation}
In the case of a barotropic fluid, we can derive the analytic expression of the sonic point.
For the case of an exponential density profile, however, the sonic point equation must be solved numerically to obtain the sonic point solution. The addition of the charge $Q$ instead, modifies the metric function as follows $f(x)=g(x)=1-2M/\sqrt{x^2+\ell^2}+Q^2/(x^2+\ell^2)$, which recovers the Reissner-Nordström solution at $\ell=0$. 

\section{Numerical results}\label{sec:results}
In this section, we present the numerical results obtained for regular black holes and Simpson-Visser spacetimes. We will examine all the different fluid models described above and present plots of the luminosity profiles to illustrate the differences between the two models and the effect of the different fluids on this observable.
\subsection{RBH solutions}

\begin{figure}[t]
\includegraphics[width=0.41\linewidth]{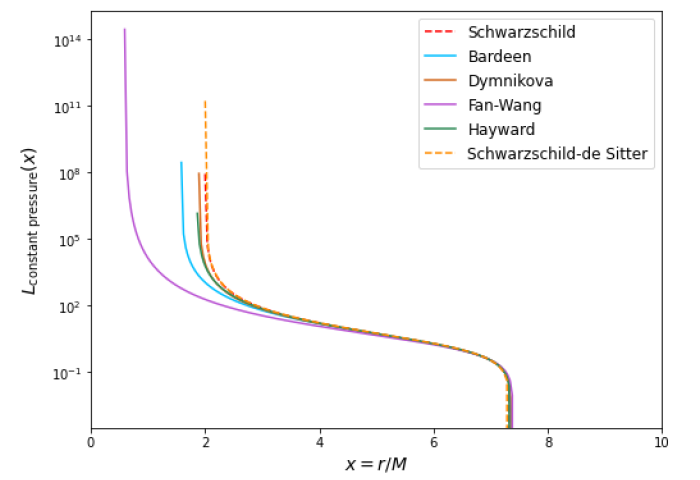}\hspace{2cm}\includegraphics[width=0.4\linewidth]{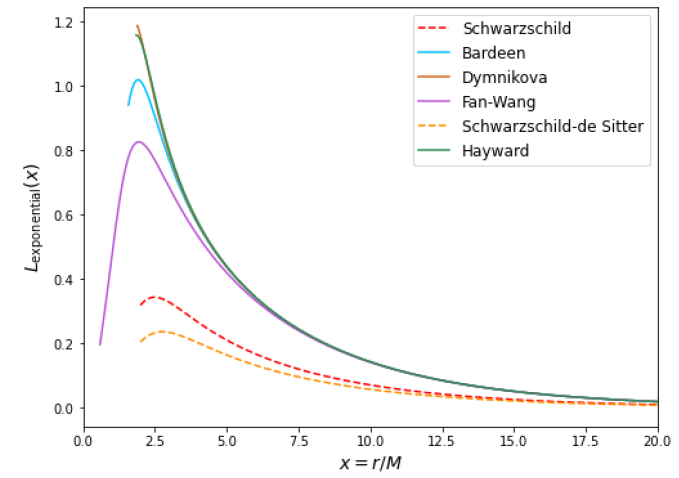}

\includegraphics[width=0.4\linewidth]{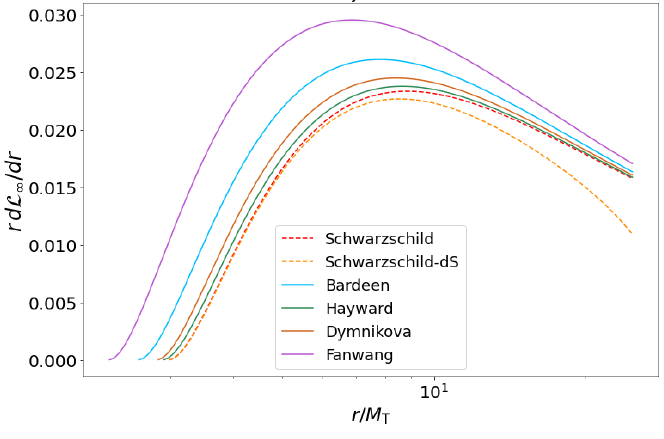}\hspace{2cm}\includegraphics[width=0.41\linewidth]{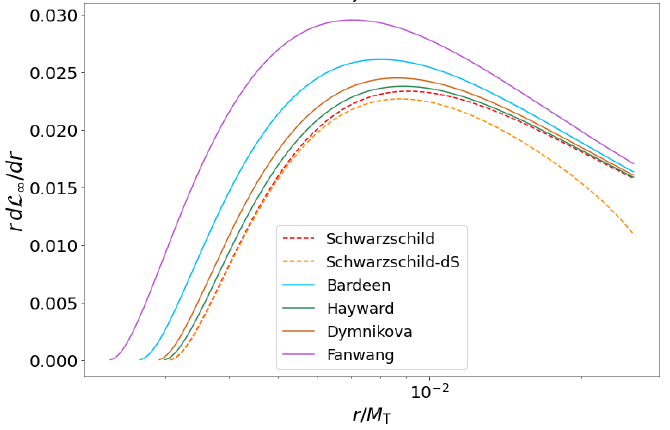}
\caption{Composite $2\times2$ of luminosity profiles for RBH solutions ($M=1$\,AU).  \textit{Upper panels}: Bondi luminosity in logarithmic scale, $\log_{10}[L(x)]$, as a function of $x=r/M$. \textit{Upper left} (dark fluid, $P=-0.75$, $C=1.5$, $C_{3}=50$): solutions span several orders of magnitude. The Fan-Wang trend is noticeably higher than the other solutions, which are more closely clustered together. \textit{Upper right} (exponential profile, $\rho_{0}=0.75$, $r_{0}=5$, $C_{1}=1$, $C_{2}=-1$): the Bondi luminosity shows a peak followed by a decline towards the horizon and similar to the NT profiles. \textit{Lower panels}: NT differential luminosity $r\,d\mathcal{L}_{\infty}/dr$ as a function of $r/M_T$, where $M_T=M + M_{rm EoS}$ and $M=1$\,AU. All solutions in both panels share the same bell-shaped trend, differing by no more than a factor of a few.}
\label{fig:lum_comparison}
\end{figure}

Table \ref{tab:RBH_combined} shows the critical-point and event horizon results for all RBH solutions. Starting from the dark fluid, we recall that the critical point here is a mathematical redefinition arising from the change in sign of $V^2$. The luminosity profiles exhibit the same behaviour as the accretion rate and produces profiles that increase indefinitely on a logarithmic scale as they approach the horizons. These profiles span several orders of magnitude across solutions (see the upper-left panel of Fig.\ref{fig:lum_comparison}). In particular, the Fan-Wang solution shows anomalously large values, suggesting a tension with this EoS, with respect to the other metrics.

For the exponential profile, critical point analysis shows that sonic points exist for all RBHs, but not for classical solutions (this issue may be related to incompatibility with the procedure), see Tab.\ref{tab:RBH_combined}. The luminosity profiles show a peak, followed by a decrease, as the solutions approach the horizon (see the upper-right panel of Fig.\ref{fig:lum_comparison}). We can also see a difference in the hierarchy of the solution profiles.

The luminosity profiles of the NT model (lower panels of Fig.\ref{fig:lum_comparison}) are almost identical for all solutions and both fluid models, differing by no more than a few percent. The behaviour of the solutions is in contrast to Bondi accretion. In fact, as pointed out in Ref. \refcite{Capozziello:2025}, The Bondi model provides a more effective probe of the spacetime geometry, particularly with regard to the exponential density profile. Consequently, it is far more sensitive than the NT model to variations in the metric and the fluid.

\begin{table}[t]
\caption{Numerical results for RBH solutions
for constant pressure and exponential density profiles ($M=1$\,AU). Dark fluid parameters: $P=-0.75$, $C=1.5$, $C_{3}=50$, $a=0.5$, $q_{B}=0.65$, $q_{D}=0.452$, $\Lambda=0.001$, $l_{\rm FW}=8/27$. Exponential profile parameters: $\rho_{0}=0.75$, $r_{0}=5$, $C_{1}=1$, $C_{2}=-1$.}
\label{tab:RBH_combined}
\begin{center}
\begin{tabular}{lccccc}
\hline
 & & \multicolumn{2}{c}{\textbf{Dark fluid}} & \multicolumn{2}{c}{\textbf{Exponential profile}} \\
\cline{3-4}\cline{5-6}
\textbf{Solution} & $r_{\rm EH}$ &  $r_{\rm crit}^{\rm DF}$ & &  $r_{c}^{\rm exp}$ & $u_{c}^{\rm exp}$ \\
\hline
Schwarzschild      & 2.00       & 7.36 & & $-$   & $-$    \\
Schwarzschild-dS   & 2.00-53.74 & 7.34 & & $-$   & $-$    \\
Hayward            & 1.85       & 7.36 & &  2.18 & $-0.43$\\
Bardeen            & 1.58       & 7.36 & &  2.59 & $-0.33$\\
Dymnikova          & 1.89       & 7.36 & & 14.60 & $-0.11$\\
Fan-Wang           & 0.59       & 7.39 & & 15.09 & $-0.10$\\
\hline
\end{tabular}
\end{center}
\end{table}

\subsection{Simpson-Visser spacetime}

\begin{figure}[t]
\includegraphics[width=0.4\linewidth]{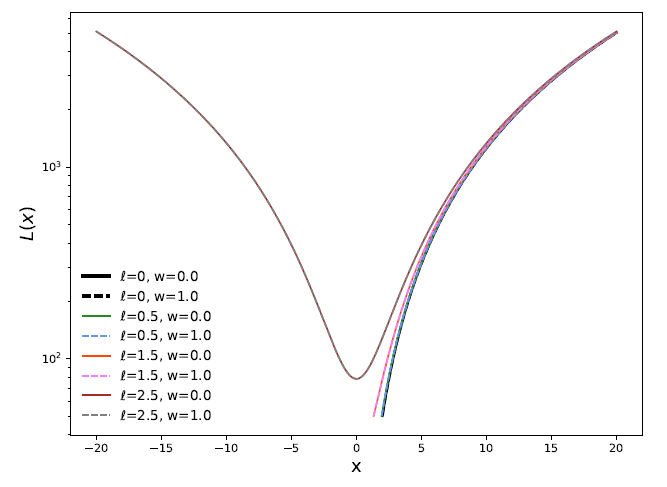}\hspace{2cm}\includegraphics[width=0.4\linewidth]{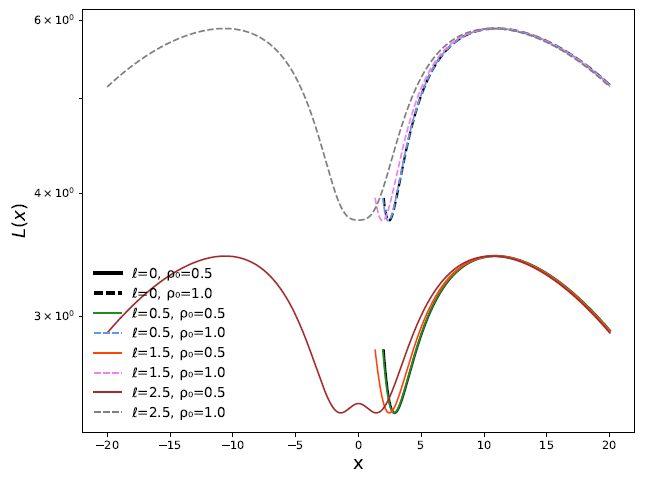}

\includegraphics[width=0.4\linewidth]{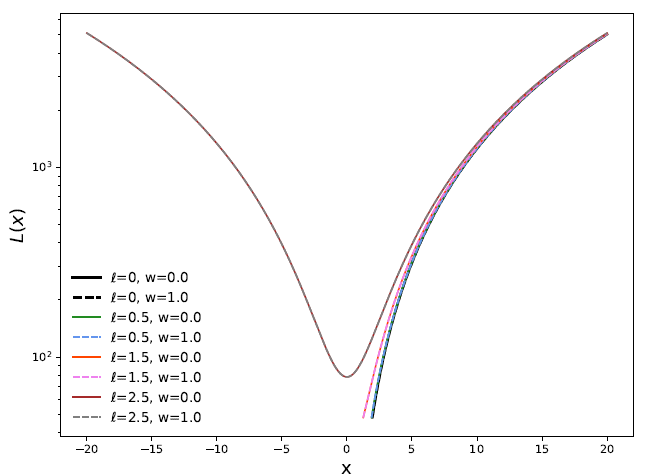}\hspace{2cm}\includegraphics[width=0.4\linewidth]{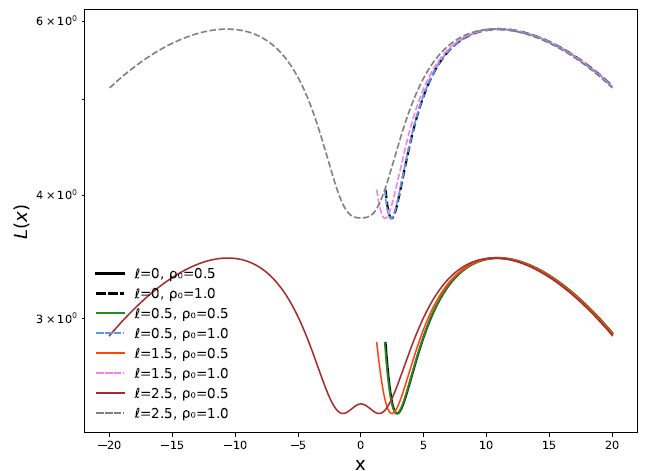}
\caption{Bondi luminosity $L(x)$ as a function of $x$ in the SV spacetime ($M=1$\,AU, $\eta=0.1$). \textit{Upper panels}: neutral SV ($Q=0$). \textit{Lower panels}: charged SV ($Q=0.3$). \textit{Left column} (barotropic fluid, $C_1=10$, $C_3=1.9$): luminosity profiles for $w=0$ and $w=1$, from $\ell=0$ to $\ell=2.5$, with wormhole profiles extended symmetrically to negative $x$. Increasing the value of $\ell$ results in a slower decline of the profiles as they move inward towards the accretion disc. The solutions remain nearly superposed with only a few exceptions. \textit{Right column} (exponential profile, $C_4=2.1$, $r_0=10$\,AU, $\rho_{0}=0.5$ and $1.0$\,AU$^{-2}$) Here, the luminosity exhibits a local minimum, followed by an increase to the horizon or throat. The inner region demonstrates the greatest sensitivity to both $\ell$ and the EoS, and thus to $\rho_0$. In all panels, the charged and neutral profiles are very similar.}
\label{fig:SV_lum}
\end{figure}

Table \ref{tab:SV_combined} shows the results for the neutral and charged SV spacetimes. For the barotropic fluid case, the solutions split into two groups according to $w$. Stiff matter ($w=1$) has a larger critical radius than dust ($w=0$) matter. Different values of $\ell$ produce nearly superposed profiles at large $x$, diverging only in the inner region. The exponential density profile is more sensitive to $\ell$ and $\rho_0$ near the horizon, showing a non-monotonic luminosity with a local minimum followed by a secondary rise towards the horizon or throat (right panels of Fig.\ref{fig:SV_lum}). However, in this second case, the differences in the critical radii values for different $\rho_0$ are smaller than for the barotropic fluid case. 

The introduction of the electric charge $Q=0.3$ produces shifts in $r_{\rm EH}$ and $x_c$ of at most $\sim3\%$. However, there are no visible effects on the luminosity profiles, as the neutral and charged curves are essentially superposed (see lower panels of Fig.\ref{fig:SV_lum}).

The wormhole configuration ($\ell=2.5$), shows a qualitatively distinct trend from the other profiles in both the neutral and charged SV cases: the profiles extend symmetrically to negative $x$. In the case of the barotropic fluid, there is a pronounced central minimum at the throat, which is not present in the other BH geometries. However, in the exponential density profile, a local maximum can be observed near the throat, followed by symmetric behaviour for negative values of $x$.

\begin{table}[t]
\caption{Sonic-point analysis for neutral ($Q=0$) and charged ($Q=0.3$) SV spacetimes. \textit{Barotropic fluid}: $C_{1}=10$, $C_{3}=1.9$, $M=1$\,AU, $\eta=0.1$. \textit{Exponential profile}: $C_{4}=2.1$, $r_{0}=10$\,AU, $\eta=0.1$. The columns $u_{c}$ and $C_{4}/C_{3}$ (the former for the neutral case and the latter for the charged case) are identical in the two cases. WH denotes the wormhole regime and therefore we do not have an event horizon.}
\label{tab:SV_combined}
\begin{center}
\setlength{\tabcolsep}{4pt}
\begin{tabular}{ccccccccc}
\hline
 & & \multicolumn{2}{c}{\textbf{Neutral} ($Q=0$)} &
   \multicolumn{2}{c}{\textbf{Charged} ($Q=0.3$)} & & \\
\cline{3-4}\cline{5-6}
$\ell$ & fluid / param. &
  $r_{\rm EH}$ & $x_{c}$ &
  $r_{\rm EH}$ & $x_{c}$ &
  $u_{c}$ & $C_{4}/C_{3}$ \\
\hline
\multicolumn{8}{l}{\textit{Barotropic fluid}} \\
0        & $w=1$ & 2.00 & 16.13 & 1.95 & 16.16 & $-0.47$ & 2.09 \\
0.5      & $w=1$ & 1.94 & 14.57 & 1.89 & 14.60 & $-0.49$ & 2.10 \\
1.5      & $w=1$ & 1.32 & 14.50 & 1.25 & 14.53 & $-0.49$ & 2.10 \\
2.5 (WH) & $w=1$ & $-$  & 14.36 & $-$  & 14.39 & $-0.49$ & 2.10 \\
0        & $w=0$ & 2.00 & 16.13 & 1.95 & 16.16 & $-0.47$ & 1.05 \\
0.5      & $w=0$ & 1.94 &  6.15 & 1.89 &  6.18 & $-0.75$ & 1.12 \\
1.5      & $w=0$ & 1.32 &  5.98 & 1.25 &  6.01 & $-0.75$ & 1.12 \\
2.5 (WH) & $w=0$ & $-$  &  5.64 & $-$  &  5.67 & $-0.75$ & 1.12 \\
\multicolumn{8}{l}{\textit{Exponential profile}} \\
0        & $\rho_{0}=0.5$ & 2.00 & 4.40 & 1.95 & 4.44 & $-0.34$ & 2.10 \\
0.5      & $\rho_{0}=0.5$ & 1.94 & 4.37 & 1.89 & 4.41 & $-0.34$ & 2.10 \\
1.5      & $\rho_{0}=0.5$ & 1.32 & 4.13 & 1.25 & 4.18 & $-0.34$ & 2.10 \\
2.5 (WH) & $\rho_{0}=0.5$ & $-$  & 4.40 & $-$  & 3.67 & $-0.34$ & 2.10 \\
0        & $\rho_{0}=1.0$ & 2.00 & 3.26 & 1.95 & 3.29 & $-0.39$ & 3.00 \\
0.5      & $\rho_{0}=1.0$ & 1.94 & 3.22 & 1.89 & 3.26 & $-0.39$ & 3.00 \\
1.5      & $\rho_{0}=1.0$ & 1.32 & 2.90 & 1.25 & 2.93 & $-0.39$ & 3.00 \\
2.5 (WH) & $\rho_{0}=1.0$ & $-$  & 3.26 & $-$  & 2.15 & $-0.39$ & 3.00 \\
\hline
\end{tabular}
\end{center}
\end{table}

\section{Conclusions and perspectives}
\label{sec:conc}

In this work, we have shown how Bondi accretion is more sensitive to spacetime geometry and the EoS of the fluid than the NT accretion model. This makes the Bondi accretion model a better discriminant for distinguishing RBHs from classical solutions\cite{Capozziello:2025}. For a dark fluid, the standard sonic-point analysis must be modified due to the vanishing sound speed. For an exponential profile, critical points are well-defined for all RBHs, but not for Schwarzschild or Schwarzschild-de Sitter geometries. When applied to the SV family\cite{Gambino:2025}, the Bondi model reveals that the location of the critical point and the luminosity profiles at small radius are primarily controlled by the parameter $\ell$, while the electric charge acts as a subleading perturbation ($\lesssim3\%$). Wormhole configurations are topologically distinct. Not only do they exhibit a luminosity minimum at the throat ($x=0$), which cannot be reproduced by any BH geometry, but they also show the greatest deviations from classical solutions, reaching almost $\sim30\%$.

Several natural extensions of this work are currently under investigation, such as generalisation to rotating spacetimes for comparison with EHT observations, for which spin estimates are available. A second direction involves the matter sector: alternative dark matter distributions may lead to different predictions in terms of both sonic radius and emitted luminosity \cite{Luongo:2025iqq}. Instead, the results obtained for both neutral and charged SV solutions motivated the study of additional hairy parameters, magnetically charged configurations, particle productions \cite{Belfiglio:2022qai} and nonlinear electrodynamics couplings and repulsive gravitational effects arising in the exterior regions of regular spacetimes\cite{Boshkayev:2023,Luongo:2014qoa}. On the observational side, the numerical signatures identified here may become accessible with future instruments. GRMHD simulations, combined with next-generation imaging by the EHT, could help determine whether the phenomenology of the accretion of astrophysical compact objects is more accurately described by regular or classical BH spacetimes.
\section{Acknowledgments}

S.G. thanks S. Capozziello, R. Giamb\`{o} and O. Luongo for scientific support and collaboration on \refcite{Capozziello:2025,Gambino:2025}.

\bibliographystyle{ws-ijmpa}
\bibliography{sample}
\end{document}